\documentclass[iop,tighten,apj]{emulateapj}

\slugcomment{ApJL accepted}

\begin{document}

\title{Imaging of the CCS $22.3$~GHz emission in the Taurus Molecular Cloud complex}

\shorttitle{Imaging of the CCS emission in the TMC}

\author{Nirupam Roy\altaffilmark{1}, Abhirup Datta\altaffilmark{2}, Emmanuel Momjian\altaffilmark{1}, and Anuj P. Sarma\altaffilmark{3}}
\altaffiltext{1}{National Radio Astronomy Observatory, P. O. Box O, 1003 Lopezville Road, Socorro, NM 87801, USA}
\altaffiltext{2}{Center for Astrophysics \& Space Astronomy, 593 UCB, University of Colorado Boulder, Boulder, CO 80309, USA}
\altaffiltext{3}{Physics Department, DePaul University, 2219 N. Kenmore Ave., Chicago, IL 60614, USA}

\shortauthors{N. Roy et al.}

\begin{abstract}
Thioxoethenylidene (CCS) is an abundant interstellar molecule, and a good 
tracer of high density and evolutionary stage of dense molecular clouds. It is 
also a suitable candidate for Zeeman splitting observations for its high 
splitting factor and narrow thermal linewidths. We report here EVLA $22.3$~GHz 
observations of three dense molecular cores TMC-1, TMC-1C and L1521B in the 
Taurus Molecular Cloud complex to image the CCS $2_1-1_0$ transition. For all 
three sources, the clumpy CCS emission is most likely tracing the starless 
cores. However, these compact structures account for only $\sim1-13\%$ of the 
integrated emission detected in single-dish observations, indicating the 
presence of significant large scale diffuse emission in favorable conditions 
for producing CCS. 

\end{abstract}

\keywords{ISM: general --- ISM: individual objects (L1521B, TMC-1, TMC-1C) --- ISM: molecules --- radio lines: ISM}

\section{Introduction}
\label{sec:intro}

Understanding the process of star formation is one of the challenging problems 
of modern astrophysics. In spite of significant progress in both theoretical 
and observational studies of the subject, there are unsolved questions about 
the driving and regulatory mechanisms of star formation. To critically test 
different star formation theories, it is important to have detailed knowledge 
of physical properties, e.g. structure and kinematics, temperature, density, 
and chemical abundances of different phases of molecular clouds and star 
forming cores. A good tracer of high density molecular clouds is the 
thioxoethenylidene (CCS) molecule, which has transitions with intrinsically 
narrow thermal linewidths. Therefore, observation of CCS can be utilized to 
probe physical properties and to better understand the star formation process.

With the $45$~m telescope of Nobeyama Radio Observatory, \citet{suz84} and 
\citet{kai87} were the first to detect unknown spectral lines from 
astronomical sources at $45.4$ and $22.3$~GHz, which were later identified by 
laboratory microwave spectroscopy to be the CCS $4_3-3_2$ and $2_1-1_0$ 
transitions, respectively \citep{sai87}. Chemical evolution models 
\citep{aik01,aik05} along with further single-dish observations 
\citep[e.g.][]{suz92,lai00} of starless cores reveal that CCS is abundant in 
the initial phase of molecular gas (before the onset of gravitational 
collapse). For a discussion on the formation pathways of CCS and its impact on 
astrochemistry, see \citet[][and references therein]{sak07}. CCS transitions 
have moderate optical depth and very narrow thermal linewidths with no 
hyperfine splittings. All these characteristics make CCS an ideal probe of the 
structure and dynamics of dark molecular clouds.

To date, most of the observational studies of CCS were done using single-dish 
radio telescopes \citep[e.g.][]{fue90,suz92,sca96,wol97,ben98,hir02,hir04, 
tat10}, while radio interferometric observation of CCS is sparse. Earlier, 
\citet{vel95} and \citet{kui96} combined single-dish and the Very Large Array 
(VLA) $22.3$~GHz interferometric data to image CCS in two protostellar cores 
B335 and L1498. The Berkeley Illinois Maryland Association (BIMA) array was 
used by \citet{oha99} to study the CCS $33.7$~GHz transition in the starless 
core L1544, and by \citet{lai00} to observe 11 other molecular cores. The BIMA 
array was also used by \citet{kua94} to observe the CCS $81.5$~GHz emission 
from Sgr~B2. The VLA have been used by \citet{deg05}, \citet{hir10} and 
\citet{dev11} to observe the CCS $22.3$~GHz line in the star-forming region 
B1-IRS, the protostellar core L483, and the infrared dark cloud G$19.30+0.07$, 
respectively. 

\begin{figure*}
\begin{center}
\includegraphics[width=4.49cm,height=5.9cm,angle=-90]{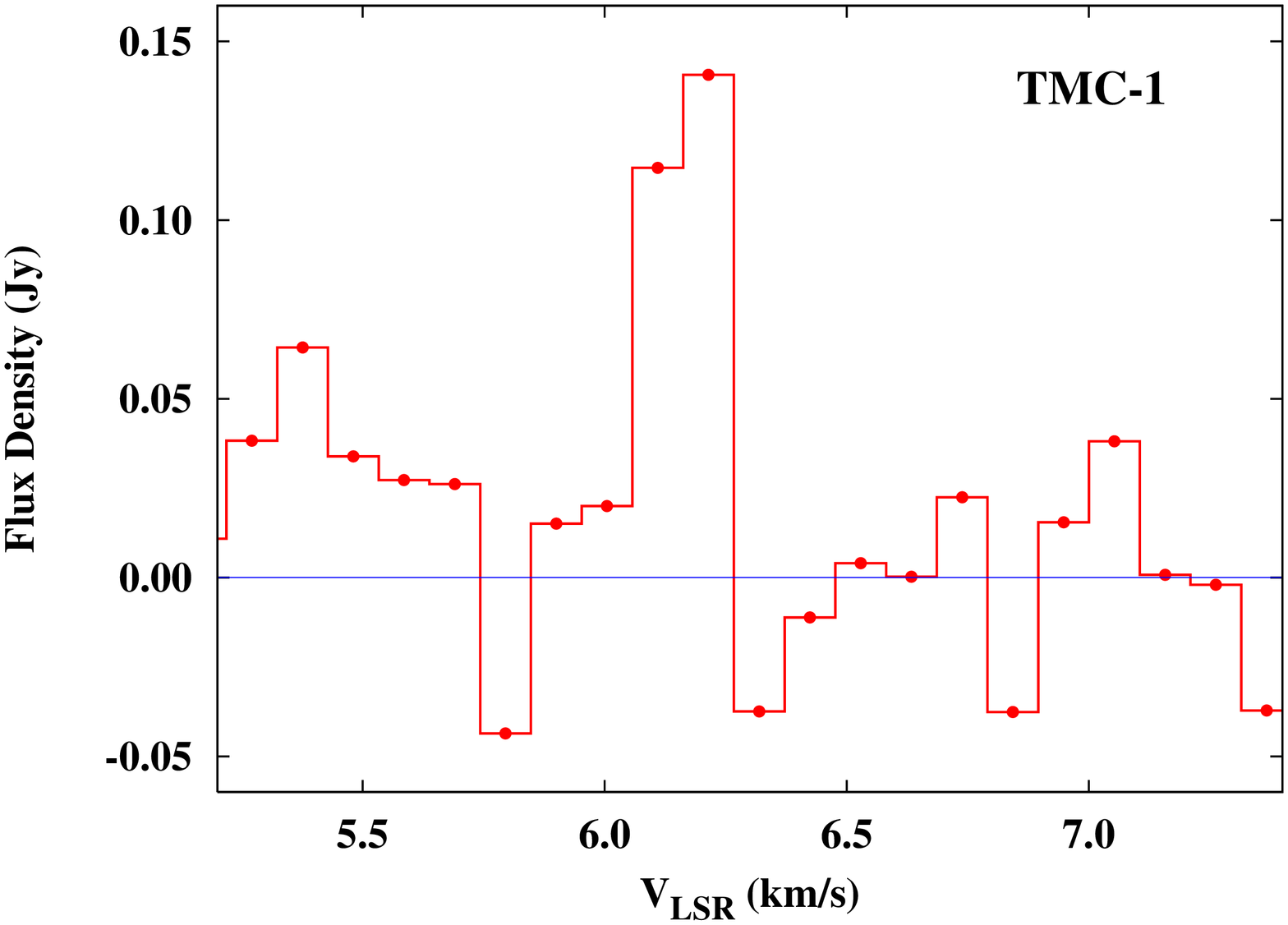}\includegraphics[width=4.49cm,height=5.9cm,angle=-90]{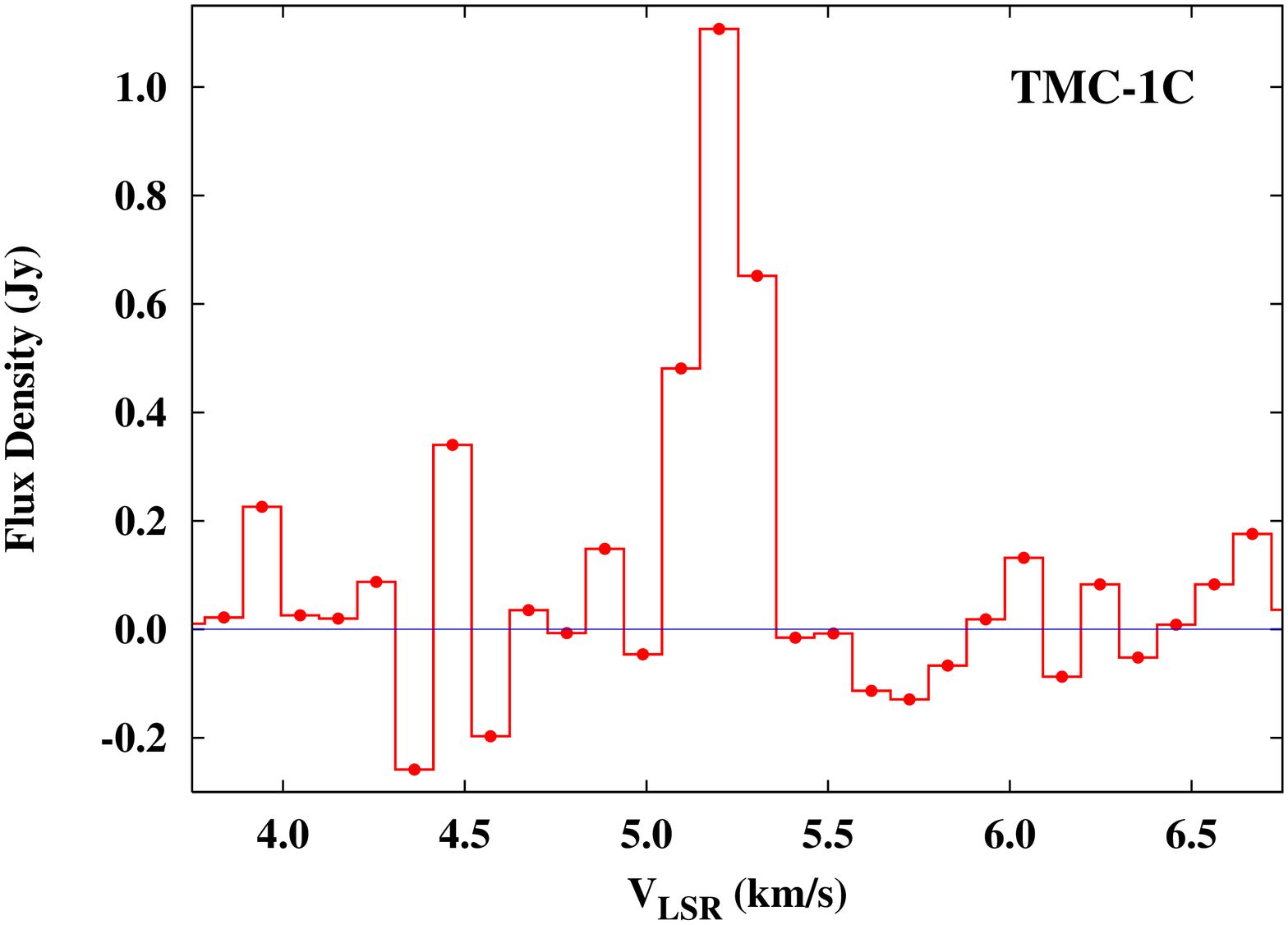}\includegraphics[width=4.49cm,height=5.9cm,angle=-90]{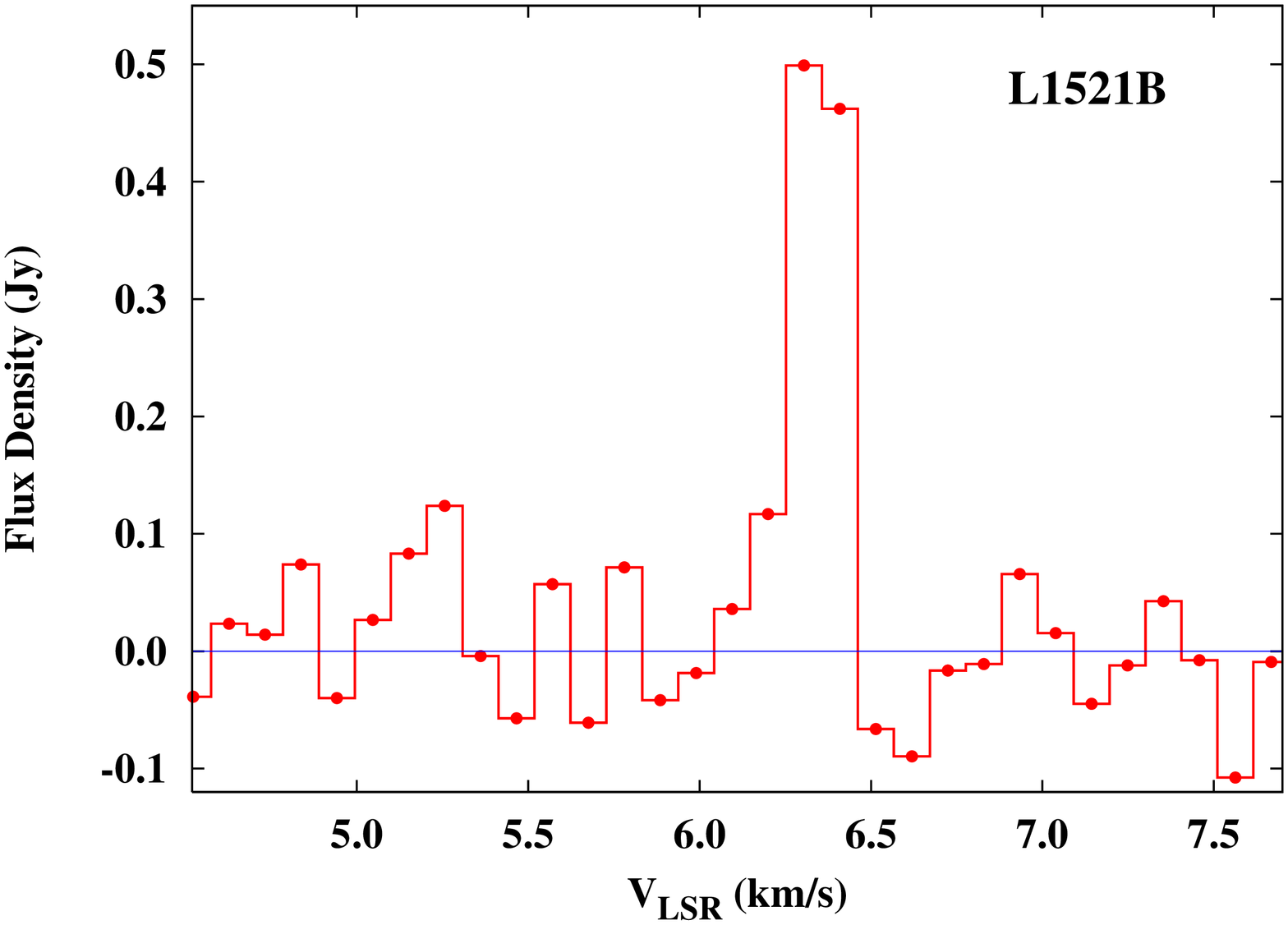}
\includegraphics[height=6.00cm,angle=-00]{fig04.ps}\includegraphics[height=6.00cm,angle=-00]{fig05.ps}\includegraphics[height=6.00cm,angle=-00]{fig06.ps}
\caption{\label{fig:allspec} {\it Top:} CCS $22.3$~GHz spectra for TMC-1, TMC-1C and L1521B from the corresponding regions shown in the bottom panels. {\it Bottom:} Integrated CCS emission images for TMC-1, TMC-1C and L1521B with a restoring beam size of $10.2\arcsec\times10.2\arcsec$. Contour levels (in Jy~beam$^{-1}$m~s$^{-1}$) are {\it Left:} TMC-1 (6.5, 13, 15, 17, 17.7), {\it Middle:} TMC-1C (5, 8, 11, 15, 19, 26, 28.5), {\it Right:} L1521B (6, 10, 13, 14.7).}
\end{center}
\end{figure*}

\begin{figure*}
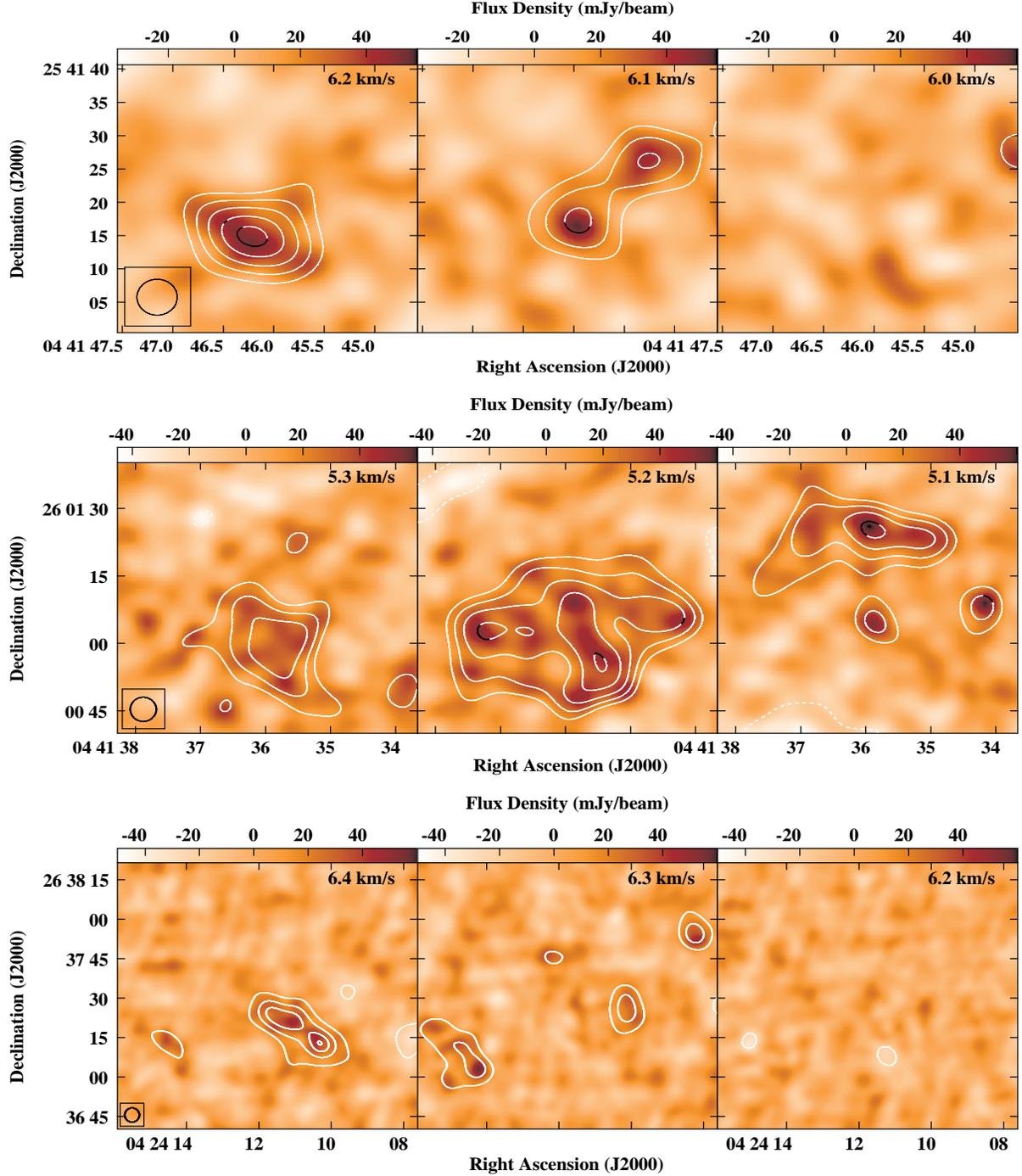

\begin{center}
\includegraphics[height=16.40cm,width=6.20cm,angle=-90]{fig07.ps}
\includegraphics[height=16.40cm,width=6.20cm,angle=-90]{fig08.ps}
\includegraphics[height=16.40cm,width=6.20cm,angle=-90]{fig09.ps}
\caption{\label{fig:allcmap} Channel maps showing CCS emission for TMC-1, TMC-1C and L1521B in the top, middle and bottom panels respectively. Color plots are $5.4\arcsec$ resolution images overlaid with corresponding contours from $10.2\arcsec$ images. Contour levels (in mJy~beam$^{-1}$) are {\it Top:} TMC-1 $14.67\times$(-3, 3, 4, 5, 6, 6.8), {\it Middle:} TMC-1C $16.67\times$(-3, 3, 4, 5, 6, 6.8), {\it Bottom:} L1521B $14.33\times$(-3, 3, 5, 6.5, 7.4).}
\end{center}
\end{figure*}

\section{Observation and Analysis}
\label{sec:ona}

The EVLA observations were carried out on October 05, 2010 during the move 
time from DnC to C array configuration (DnC~$\rightarrow$~C). The EVLA K-band 
centered at the CCS 22.3 GHz transition was used with a total bandwidth of 
$2.0$~MHz and 256 spectral channels. The resulting spectral resolution was 
$\sim 8$~kHz ($\sim 0.1$ km~s$^{-1}$). $3C~147$ was observed for flux 
calibration. The calibrator sources J$0440+2728$ (for TMC-1 and TMC-1C) and 
J$0429+2724$ (for L1521B) were used to calibrate the complex gains, and were 
observed for $1.5$ minute for every $3.5$ minute time on the target source. 
Total observation time was $2$~hours per source resulting in an on-source time 
of $\sim 1$~hour for each target.

Data analysis was carried out using the NRAO Astronomical Image Processing 
System ({\small AIPS}). After removing bad data, the flux density scale 
and instrumental phases were calibrated. Phase only calibration solutions were 
obtained first, followed by amplitude and phase calibration with longer 
solution intervals. The target source data were then split after applying the 
calibration solutions, and were used to make image cubes to check for the CCS 
emission. No continuum emission was detected in either high or low resolution 
images of the target fields. Multiscale imaging with natural weighting was 
used to make final emission line image cubes with restoring beam sizes of 
$2.0$, $5.4$ and $10.2\arcsec$. From the image cubes, moment maps were made 
with a $3\sigma$ flux cutoff. Integrated spectra for the region corresponding 
to the moment maps were extracted. The channel maps were used to study the 
structure of the emission regions. 

\section{Results}
\label{sec:result}

CCS $22.3$~GHz transition is detected in emission for all of these three 
sources. Their spectra are shown in Figure~(\ref{fig:allspec}) top panels, and 
the integrated CCS emission (``moment~0'') maps with a restoring beam size of 
$10.2\arcsec$ are shown in bottom panels. The peak line flux densities detected 
in the EVLA observations are approximately $0.15$, $1.10$ and $0.50$ Jy for 
TMC-1, TMC-1C and L1521B, corresponding to $5.4$, $4.4$, $4.5$ K respectively. 
Channel maps showing the CCS emission are presented in 
Figure~(\ref{fig:allcmap}). The color images are at $5.4\arcsec$ resolution 
overlaid on corresponding contours from the $10.2\arcsec$ resolution images. 
As expected, the emission is confined within $0.2 - 0.3$ km~s$^{-1}$ velocity 
width. Both the integrated and channel maps show prominent clumpy emission 
structures more likely tracing the highest density regions in the molecular 
cores. These structures are more clearly visible in the $5.4\arcsec$ 
resolution images where much of the extended diffuse emission is resolved out. 
For a typical excitation temperature of $5$~K \citep{suz92,dev11}, the CCS 
column densities are $(0.96 - 1.48) \times 10^{13}$~cm$^{-2}$. Assuming a 
fractional abundance of $0.9\times 10^{-10}$ for CCS \citep{lai00}, the 
resulting H$_2$ column densities are $(1.07 - 1.65) \times 10^{23}$~cm$^{-2}$. 
If the line of sight extent of the cores are similar to their extent on the 
plane of the sky, then the corresponding H$_2$ number densities will be $(2.6 
- 6.5) \times 10^{4}$~cm$^{-3}$. This is consistent with the fact that CCS is 
excited at densities in the range of $10^{4} - 10^{5}$ cm$^{-3}$ 
\citep{wol97}. We note that the CCS emission may come either the dense 
starless cores in the early stages of star formation, or the core envelopes in 
the later stages \citep{lai00}. For these sources, we do not see any envelope 
or ring-like structures in CCS emission, suggesting that the clumps are 
actually tracing starless cores.

We also note that, using a conversion factor of $2.8$ Jy~K$^{-1}$ \citep{tak10} 
for the Nobeyama $45$~m telescope, the single-dish integrated flux densities 
of the $22.3$~GHz emission are $2.94$, $1.93$ and $1.71$ Jy\,km\,s$^{-1}$ for 
TMC-1, TMC-1C and L1521B respectively \citep{suz92}. This implies that only 
$\sim1-13\%$ of the emission is in the compact structures seen with the EVLA. 
The rest of the emission that is resolved out by the EVLA is most likely 
arising from large scale structures greater than $\sim 45\arcsec$ 
(corresponding to the the shortest baseline of $\sim 2.6$~k$\lambda$ of 
DnC~$\rightarrow$~C array observations). For example, the source L1521B was 
mapped with a single-dish \citep{hir04}, and found to have large scale 
emission over an area of $\sim 7\arcmin\times4\arcmin$, which is several times 
larger than the shortest spacing of our observation.

\section{Discussions}
\label{sec:discn}

Apart from being a tracer of high density, the CCS molecule is a good 
indicator of age for dense molecular clouds. It is now well-established that 
molecular clouds in different stages of the pre-stellar evolution have 
different molecular abundances. Younger cores are rich in carbon-chain 
molecules such as CCS and HC$_3$N, while more evolved cores (close to the 
protostellar formation via gravitational collapse) are rich in late-type 
molecules such as NH$_3$, N$_2$H$^+$ and CN \citep{suz92,cc05,hb10}. 
Therefore, the ratio of the column densities of NH$_3$ and CCS is a good 
tracer for the age of the dark clouds. Similarly, the C$^{18}$O to CCS column 
density ratio indicates the production efficiency of CCS-like carbon-chain 
molecules in different physical conditions.

Based on the Nobeyama $45$~m telescope observations, $N_{NH_3}/N_{CCS}$ is 
$1.7$, $2.9$, and $18.4$ for L1521B, TMC-1 and TMC-1C, respectively 
\citep{suz92}, showing that they are in different phases between young ``dense 
core'' and ``core with star formation''. Ordering by the age, L1521B is in an 
early phase, TMC-1 is moderately old and TMC-1C is in a late phase of the 
pre-stellar evolution. The C$^{18}$O to CCS column density ratio is $83$, 
$50$, $102$ for L1521B, TMC-1 and TMC-1C respectively 
\citep{ben83,suz92,hir04}. Interestingly, the fraction of CCS single-dish flux 
detected with EVLA in the clumpy structures increases with increasing 
$N_{C^{18}O}/N_{CCS}$ ratio for these lines of sight, but it does not 
correlate with the age of the cores. For high $N_{C^{18}O}/N_{CCS}$, the CCS 
emission is slightly clumpy, whereas for low $N_{C^{18}O}/N_{CCS}$, i.e. in 
favorable conditions of CCS production, the emission is more diffuse. This may 
be indicative of the fact that in favorable conditions, CCS is formed on and 
spread over large diffuse regions. Future observations of more sources with 
the EVLA, simultaneously covering NH$_3$ and CCS transitions, may be useful to 
check the statistical significance of this trend.

The other reason that makes CCS an interesting molecule is the high Zeeman 
splitting factor of its GHz transitions \citep{shi00}. For example, the Zeeman 
splitting factor of CCS $2_1-1_0$ $22.3$~GHz transition is $0.767$ 
Hz~$\mu$Gauss$^{-1}$. This, along with its intrinsically narrow thermal 
linewidths, make CCS a suitable candidate to determine the line of sight 
component of the local magnetic field in dense molecular clouds, particularly 
in preprotostellar cores. Systematic measurement of the magnetic field 
strengths and the mass-to-flux ratios in molecular clouds at different phases 
of the star formation process and at different densities may play a crucial 
role in constraining the theories of star formation. So far, with single-dish 
observations, \citet{shi99} have reported a marginal detection of Zeeman 
splitting of CCS $4_3-3_2$ transition, and \citet{lev01} have reported a 
tentative detection of Zeeman splitting of CCS $2_1-1_0$ transition. The 
improved sensitivity and capabilities of the EVLA at high frequencies have now 
opened the opportunity of carrying out sensitive interferometric observation 
for the thermal line Zeeman splitting using CCS transitions. From dust 
polarization measurements, the magnetic field towards starless cores in the 
TMC is estimated to be $10 - 160 \mu$G \citep{cru04,kir06}. With the present 
EVLA sensitivity, the median magnetic field of $\sim 80 \mu$G would be 
detected at $5\sigma$ level in 36 hours of on-source time, whereas only 9 
hours is sufficient to detect the strongest magnetic fields ($\sim 160 \mu$G) 
for TMC-1C using the Zeeman splitting of the CCS $22.3$ GHz transition. In 
this context, it is important to first identify suitable lines of sight with 
strong CCS emission and adequate flux in compact components for Zeeman 
splitting observations. In the near future, we plan to carry out further 
observations of candidate sources to identify possible targets for a CCS 
Zeeman splitting survey with the EVLA. 

\section{Conclusions}
\label{sec:concl}

EVLA observations of three dense molecular cores TMC-1, TMC-1C and L1521B in 
the Taurus Molecular Cloud complex were carried out to image CCS emission from 
these regions. CCS $2_1-1_0$ $22.3$~GHz narrow emission line is detected for 
all of these three sources. The emission is found to have prominent clumpy 
structures most likely tracing the high density regions in the starless 
cores. Only $\sim 1 - 13\%$ of the integrated emission detected in previous 
single-dish observations comes from these compact emission structures. This 
indicates that most of the CCS emission comes from low surface density 
extended structures. The unresolved fraction of CCS is correlated with 
$N_{C^{18}O}/N_{CCS}$ ratio for these three lines of sight, suggesting more 
extended distribution CCS in favorable conditions for producing CCS. Such 
exploratory observations to image CCS in dense clouds are important to 
identify suitable target sources for future EVLA Zeeman splitting observations.

\acknowledgments

We are grateful to Jayaram N. Chengalur for helpful discussions and many useful comments on this paper. NR is a Jansky Fellow of the NRAO. AD is supported by NASA through NASA Postdoctoral Program at the University of Colorado Boulder, administered by ORAU through a contract with NASA. We are also grateful to the anonymous referee for prompting us into substantially improving this paper.

{\it Facilities:} \facility{EVLA}\\

Disclaimer: This is an author-created, un-copyedited version of an article accepted for publication in the Astrophysical Journal Letters. IOP Publishing Ltd is not responsible for any errors or omissions in this version of the manuscript or any version derived from it. The definitive publisher authenticated version will be available online at {\tt http://iopscience.iop.org/}.

\end{document}